\documentclass[12pt]{article}

\usepackage{amssymb,amsmath,amsfonts,eurosym,geometry,ulem,graphicx,caption,color,setspace,sectsty,comment,footmisc,caption,natbib,lscape,subfigure,array,threeparttable}
\usepackage[noblocks]{authblk}
\usepackage{hyperref}
\hypersetup{
    citecolor=blue,
   colorlinks=true,
   linkcolor=blue,
   filecolor=magenta,
   urlcolor=cyan,
}
\newcolumntype{L}[1]{>{\raggedright\let\newline\\arraybackslash\hspace{0pt}}m{#1}}
\newcolumntype{C}[1]{>{\centering\let\newline\\arraybackslash\hspace{0pt}}m{#1}}
\newcolumntype{R}[1]{>{\raggedleft\let\newline\\arraybackslash\hspace{0pt}}m{#1}}

\begin{document}

\begin{titlepage}
\title{Identifying Opportunities to Improve the Network of Immigration Legal Services Providers\thanks{We thank members of the Immigration Policy Lab for comments, Aunt Bertha for sharing data, and Rachel Kim, Niki Ngyuen, Jong Beom Lim, Manasa Kumarappan, Sean Lee, Leila Braganza, and Phoebe Quinton for research assistance. We recognize funding from the Charles Koch Foundation. Funders had no role in the data collection, analysis, decision to publish, or preparation of the manuscript.}}

\author[a,b,1]{Vasil Yasenov} 
\author[a,c]{David Hausman} 
\author[a]{Michael Hotard} 
\author[a]{Duncan Lawrence} 
\author[a,d]{Alexandra Siegel} 
\author[a]{Jessica S. Wolff}
\author[a,e]{David D. Laitin} 
\author[a,e,1,2]{Jens Hainmueller}
\affil[a]{Immigration Policy Lab, Stanford University}
\affil[b]{IZA Institute of Labor Economics}
\affil[c]{Lane Center for the American West, Stanford University}
\affil[d]{Department of Political Science, University of Colorado Boulder}
\affil[e]{Department of Political Science, Stanford University}


\date{\today}
\maketitle
\clearpage
\begin{abstract}

\singlespacing
Immigration legal services providers (ISPs) are a principal source of support for low-income immigrants seeking immigration benefits. Yet there is scant quantitative evidence on the prevalence and geographic distribution of ISPs in the United States. To fill this gap, we construct a comprehensive, nationwide database of 2,138 geocoded ISP offices that offer low- or no-cost legal services to low-income immigrants. We use spatial optimization methods to analyze the geographic network of ISPs and measure ISPs' proximity to the low-income immigrant population. Because both ISPs and immigrants are highly concentrated in major urban areas, most low-income immigrants live close to an ISP. However, we also find a sizable fraction of low-income immigrants in underserved areas, which are primarily in midsize cities in the South. This reflects both a general skew in non-governmental organization service provision and the more recent arrival of immigrants in these largely Southern destinations. Finally, our optimization analysis suggests significant gains from placing new ISPs in underserved areas to maximize the number of low-income immigrants who live near an ISP. Overall, our results provide vital information to immigrants, funders, and policymakers about the current state of the ISP network and opportunities to improve it. \\

\vspace{.5cm}

\bigskip
\end{abstract}
\thispagestyle{empty}
\end{titlepage}

\doublespacing


\setcounter{page}{1}

\section{Introduction}
Immigrants often navigate legal challenges most citizens never encounter, such as renewing visas, petitioning for family reunification, pursuing naturalization, or defending themselves against deportation in immigration courts. Navigating these complex bureaucratic and legal procedures competently requires assistance from legal professionals. Demand for immigrant legal assistance is higher than ever given the growing immigrant population and recent policy developments, such as half a million new deportation proceedings in 2019 \citep{trac}, changes to public charge rules that are affecting eligibility for permanent residency, and the proposed cancellation of various programs (e.g., Deferred Action for Childhood Arrivals and Temporary Protected Status) involving hundreds of thousands of immigrants. Yet the federal government provides no legal services directed to the needs of the immigrant population; its Legal Services Corporation, the nonprofit organization that administers federal funding for legal services, excludes immigration legal services \citep{Heeren_2011}. Indeed, in 1996, Congress barred organizations that take federal funding from using their non-federal funds to represent most immigrant clients \citep{Heeren_2011}. 

In response, nonprofit organizations, which we call immigration legal services providers (ISPs), have filled the void. ISPs provide low- or no-cost legal services for immigrants and refugees. They include charities, ethnic associations, humanitarian organizations, churches, and law clinics. Even though ISPs have become a principal source of support for low-income immigrants facing legal issues, we have scant quantitative evidence on their prevalence and geographic distribution across the United States. While research has highlighted the important role ISPs can play in supporting low-income immigrants \cite{degraauwMaking2016,cortes_counting_1998,Brown_Philanthropic_2014,wong2016does,yasenov2019standardizing,ryo2019beyond,chand_serving_2020}, extant work typically has focused on specific ethnic groups and specific geographic areas and has not compiled comprehensive data on the nationwide network of ISPs. (See \cite{chand_serving_2020} for an excellent first step in this direction.) Identifying this geographic distribution of ISPs and its coverage is of fundamental importance because ISPs facilitate  access to free or low-cost legal aid. Indeed, proximity to ISPs may be critical for the integration of immigrants into their communities \citep{allard2009out,owens2005congregations,roth2016re}. At the same time, research has raised the concern that there may be a geographic mismatch between the provision of service and the immigrant population most in need \cite{roth2016re,ryo2019beyond}. 

In this study, we collect a comprehensive database of 2,138 geocoded, low- or no-cost immigrant legal service providers in the United States. We pair this new database with ZIP code-level data on the low-income immigrant population and leverage spatial optimization methods to analyze the geographic distribution of ISPs and their proximity to low-income immigrants. In particular, we answer three questions: What is the geographic distribution of ISPs? What are the most underserved, overserved, and well served areas? How can the ISP network be improved through the placement of new locations to maximize the number of low-income immigrants nearby? The answers to these questions matter for assessing and improving legal services provision for low-income immigrants.

We find that most low-income immigrants live relatively close to an ISP; 75\% of foreign-born people whose household income is below 150\% of the Federal Poverty Guidelines live within 6.6 miles of an ISP. This pattern reflects the fact that the immigrant population is highly concentrated in major urban areas, which also have dense ISP networks. Yet we also find that many low-income immigrants are not well covered by the current network. About 1.5 million low-income immigrants live in areas with no ISP within 12 miles, our criterion for access. Many of these areas with limited coverage are in midsize cities in the South. Further tests suggest that this pattern is partly driven by the fact that these areas generally have a lower presence of non-governmental service organizations and are relatively new immigrant destinations where the ISP network has lagged in development. Finally, our analysis finds that adding new ISPs in optimal locations could significantly increase the number of low-income immigrants who have access to services. In particular, adding just one more ISP optimally in each state could reduce the number of underserved low-income immigrants by about 192,600 or approximately 12\%. In addition, optimally rearranging the network of ISPs across the entire US would reduce the number of underserved low-income immigrants by about 1.1 million and close about 70\% percent of the gap in the number of people served.

Theorists in sociology often have been skeptical about the effectiveness of ISPs due to their dependence on the state \citep{taylor2002co}, their tax status (which can prohibit advocacy activities \citep{degraauwMaking2016}), and the growing strength of anti-immigrant organizations \citep{Brown_Philanthropic_2014}. Preliminary empirical research, however, is more encouraging. Even with limited data on the presence of ISPs, findings suggest that ISPs provide essential services, enabling immigrants to overcome legal hurdles in applying for naturalization \citep{yasenov2019standardizing}, seeking adjustment of status \citep{wong2016does}, fighting deportation \citep{chand_serving_2020}, obtaining release from detention \citep{ryo2019beyond}, and procuring veterans' benefits \citep{cortes_counting_1998}. Outside the legal realm, qualitative research points to ISPs' successes in teaching English, in pressing for wage increases, and in securing identity cards \citep{degraauwMaking2016}.

The lack of a comprehensive ISP database has been a persistent problem in the literature. While researchers have compiled lists of immigrant organizations using tax data, these lists have been either geographically constrained \citep{joassart-marcelli_ethnic_2018,gleeson_assessing_2013,roth2016re} or limited to specific immigrant groups \citep{cortes_counting_1998,roth2016re,hung_immigrant_2007}. Moreover, tax data undercount ISP presence by up to 50\% \citep{gleeson_assessing_2013}. The most extensive data collection that we are aware of is \citep{chand_serving_2020}, who compiled a list of 1,079 ISPs. An important contribution of our study is that, using new methods, we have identified and validated a list of nearly twice as many active ISP offices.

\section{Data and Methods}
\subsection{ISP Data}

We defined ISPs as organizations that provide low- or no-cost assistance with official immigration-related documents or forms and/or legal representation in immigration proceedings. We used three criteria. The provider must (1) be an organization; (2) qualify by the broad US Federal Title 8 standards as a representative for immigration legal services \citep{Shannon_2009}; and (3) advertise online as providing free or low-cost legal services to immigrants and/or refugees.

We drew on three sources: (1) The Immigration Advocates Network’s National Immigration Legal Services Directory \citep{ian}, where ISPs can self-register (1,153 offices); (2) the Department of Justice (DOJ) list of Recognized Organizations and Accredited Representatives Roster \citep{bia} (1,178 offices); and (3) a list of immigrant service organizations maintained by Aunt Bertha \citep{auntbertha} (2,042 offices). We geocoded all office addresses and  clustered them by location to identify distinct offices. We then did two rounds of manual coding to check the organizations' websites and added 336 ISP locations that were not listed in any of the three sources.

\subsection{Low-income Immigrant Data}

To examine proximity, we leverage counts of the low-income, foreign-born population in the 32,989 ZIP Code Tabulation Areas (ZCTAs; henceforth ZIPs) from the American Community Survey 2014–2018 5-Year sample \citep{manson50ipums}. We define ``low-income immigrants'' as all foreign-born people with household incomes lower than 150\% of the Federal Poverty Guidelines (FPG) (i.e., an annual income of \$25,860 for a household of two). We compute the ``as the crow flies'' distance between the ZIP code centroid and the closest ISP. (Immigrants' locations within ZIP codes are not disclosed.) We then aggregate to the core-based statistical area (CBSA) level as our main definition of a ``city.'' If a ZIP code falls into multiple CBSAs, we assign it to the CBSA that contains the majority of its population. When aggregating distances to the CBSA level, we weight each ZIP code by the size of its low-income immigrant population. See the Supplementary Materials (SM) for additional results: we also analyze alternative aggregations, such as commuting zones and counties, and alternative distances, such as driving distances and driving times, and find similar results.

\subsection{Identifying Underserved Areas}

We use two approaches to identify ``underserved'' areas with a high number of low-income immigrants but relatively few ISPs nearby. First, we identify CBSAs in the upper quartiles of two marginal distributions: the number of low-income immigrants and average distance to the closest ISP. Second, we compare the current number of ISPs in each CBSA to the number of ISPs implied by our optimization algorithm (described below). 

\subsection{Spatial Optimization Algorithm}

We conduct two types of spatial optimizations. First, we identify the optimal locations for new ISPs such that we maximize the number of low-income immigrants who gain access to services. We take a service area approach and define the service area of an ISP as a 12 mile radius around its location. Low-income immigrants are counted as having access to service if their ZIP code centroid is within 12 miles of the closest ISP; otherwise they are counted as underserved. The 12 miles' distance cutoff is motivated by the fact that this is about twice the average distance to the closest ISP for the low-income immigrant population. To identify the optimal placement for the next ISP, we loop through the centroids of all underserved ZIP codes and select the ISP placement where the maximum number of immigrants gain service. This assumes that ISPs can only be placed at ZIP code centroids for computational simplicity. In the SM we also use an alternative distance cutoff (15 miles) and results are similar.

Second, we identify an optimal network that maximizes the number of immigrants served by rearranging existing ISPs. This allows us to examine to what extent the spatial distribution of the existing network is suboptimal. We use the same service area approach and first generate 200,000 random ISP networks of the size of the existing ISP network by randomly sampling, without replacement, ZIP code centroids. We then select the network that maximizes the number of immigrants with access to services. We impose a capacity constraint such that the maximum number of low-income immigrants covered by any given ISP is at or below the maximum covered under the current network: 61,725 low-income, foreign-born people. We then use an iterative procedure and randomly rearrange the locations of the ISPs with the lowest number of immigrants served until no further improvements can be made. 

We conduct both optimization analyses in two ways: nationwide and state-by-state. The nationwide analysis is informative about the overall skew in the current network (i.e. across and within states). The state-by-state analysis is informative about the skew of the network within each state and also policy relevant since funding for ISPs often depends on within-state sources.

\section{Results}

\subsection{ISP Network}
\begin{figure}[tbhp]
\centering
\includegraphics[width=\linewidth]{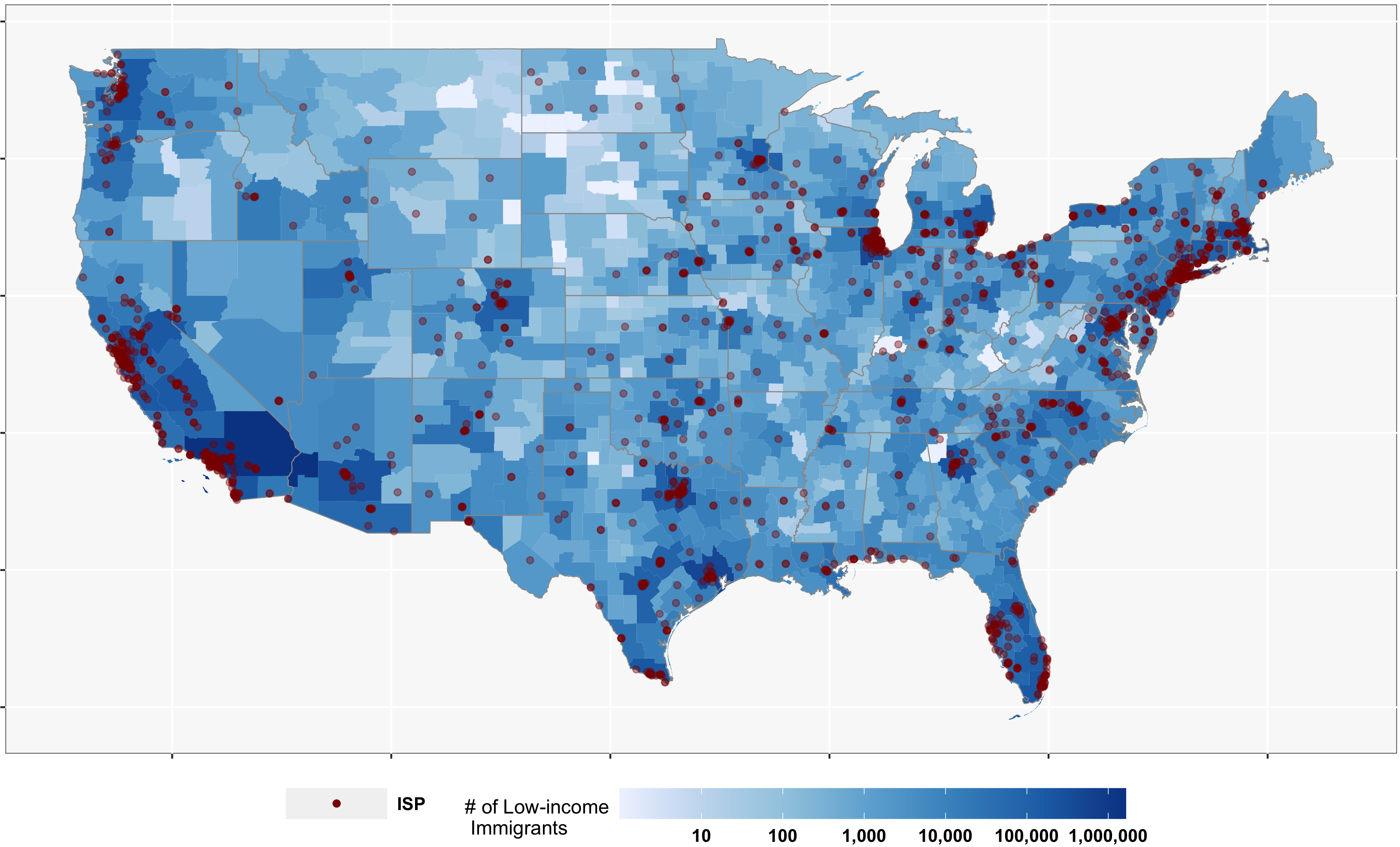}
\caption{ISP network and low-income immigrant population size across the US. Red dots indicate ISP locations. Polygons are commuting zones. }
\label{fig:map}
\end{figure}

Our final database consists of 2,138 ISP offices, 1,178 of which are registered on the Board of Immigration Appeals. Figure \ref{fig:map} presents the network of ISPs across the conterminous United States along with the size of the low-income immigrant population. Some patterns stand out. First, the ISP network is widely distributed, covering all states. ISPs cover all major urban centers, and the largest traditional immigrant-receiving hubs, such as New York City, Los Angeles, and Chicago, have some of the densest ISP networks, with 264, 124 and 100 ISPs, respectively. Although rural areas generally have low coverage, some places with small immigrant communities do have ISPs. For instance, Jamestown, ND, Riverton, WY, and Auburn, IN each host fewer than 100 low-income immigrants but contain an ISP. Second, as the concentration of red points in dark blue localities shows, both ISPs and low-income immigrants are highly concentrated in major urban areas. The correlations between these two variables range from 0.83 to 0.94, depending on the level of aggregation (e.g., county, commuting zone, CBSA, or state).

How close are low-income immigrants to an ISP? The left panel of Figure \ref{fig:deserts} presents the (empirical) quantile function of the average distance of low-income immigrants to their closest ISP. Most low-income immigrants reside relatively close to an ISP. The median low-income immigrant lives about 2.7 miles from an ISP, and 75\% of the total population of about 12 million low-income immigrants live within 6.6 miles of an office. This  pattern is driven by the fact that low-income immigrants are highly concentrated in major urban areas, which are covered by the densest ISP networks. For instance, the six largest immigrant cities---New York City, Los Angeles, Miami, Houston, Chicago, and Dallas---account for 40\% of the low-income immigrant community nationwide. The same cities account for 37.4\% of all ISPs. However, many low-income immigrants are not well covered by the current network. For example, there are about 1.5 million underserved low-income immigrants (or about 13\% of the total population) who live more than 12 miles away from the closest office, suggesting potential gains from improving the network's coverage.

\subsection{Underserved Areas}

\begin{figure}[tbhp]
\centering
\includegraphics[width=\linewidth]{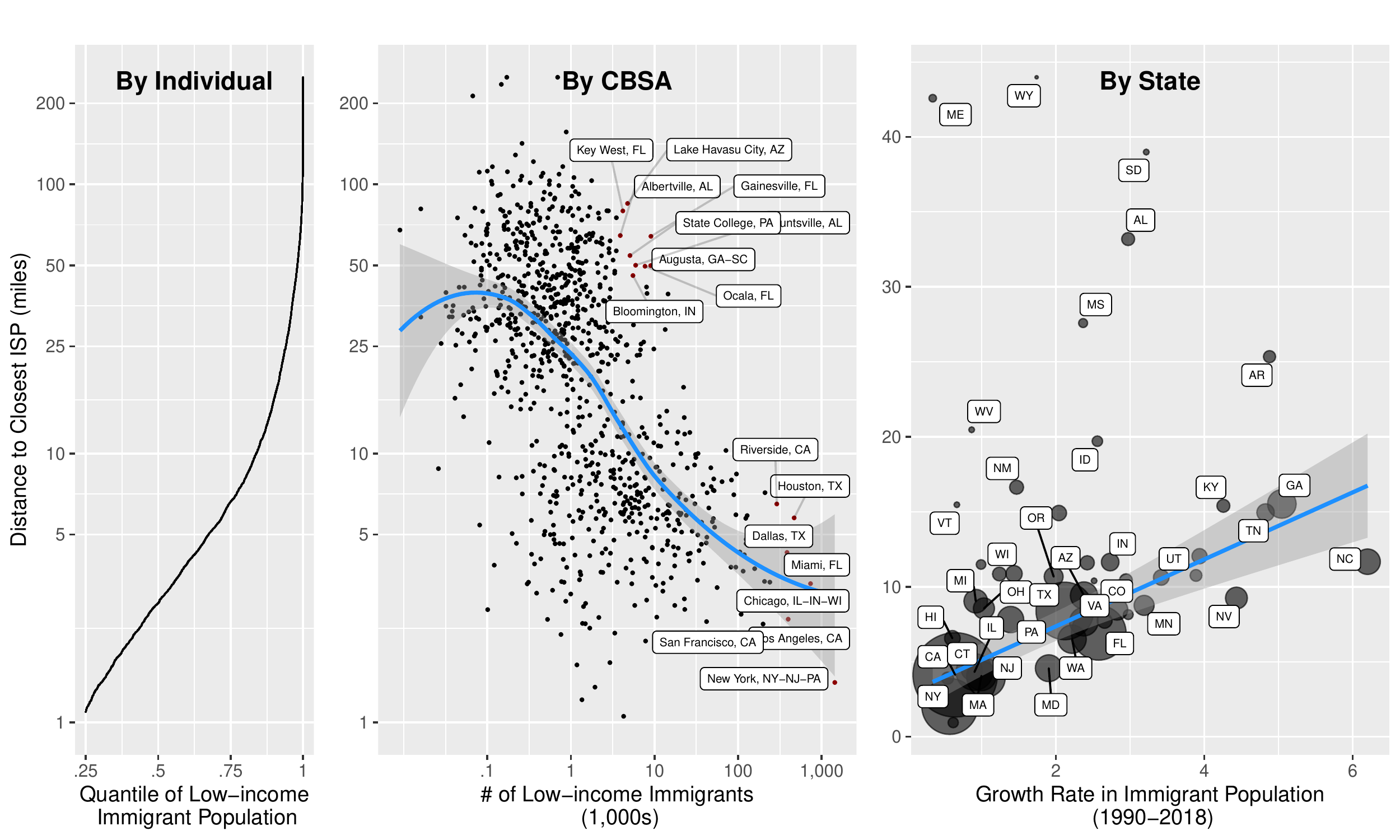}
\caption{Distance to the closest ISP at the individual (left plot), CBSA (middle) and state (right) levels. Left panel shows the fraction of low-income immigrants located within a distance to the closest ISP. Middle and right panels show distance to the ISP by size and growth of the immigrant population.} 
\label{fig:deserts}
\end{figure}

What are some of the most underserved areas? The middle panel of Figure \ref{fig:deserts} shows a CBSA-level scatter plot of low-income immigrant population size and average distance to the closest ISP. The inverse relationship indicates that, on average, cities with more low-income immigrants are better connected to the ISP network. The cities in the lower right cluster have dense ISP networks and contain the majority of the low-income immigrant population, which explains the pattern discussed above---that most immigrants live close to an ISP. The upper middle cluster of cities identifies low-service areas: cities with larger numbers of immigrants who, on average, live relatively farther from an ISP. The plot labels all cities that are above the 3rd quartile in the distributions of distance to the closest ISP (44 miles) and the size of the low-income immigrant population (3700 foreign-born people). These underserved areas are mostly in Southern midsize cities such as Augusta, GA, Huntsville, AL, and Gainesville, FL. In these areas, the median distance to the closest ISP is 60 miles, and the median number of low-income immigrants is 5562.

What might explain this clustering of underserved areas? While a full investigation of this question is beyond the scope of this study, we explore two hypotheses. The first is that the ISP network lags behind the diversification of immigrant settlement patterns, with less settlement in recent decades in traditional hubs and more in new destinations  \citep{massey2008geographic,massey2016border}. The right panel of Figure \ref{fig:deserts} provides support for this idea by comparing the state-level average distance to the closest ISP with total immigration growth between 1990 and 2018. New destination states, which have experienced higher growth in immigration over this time period (e.g., Georgia, North Carolina, and Arkansas), are also, on average, relatively farther from an ISP. This suggests that the ISP network has not caught up with the diversification of immigrant settlement patterns.

A second hypothesis is that the underserved areas for immigrant legal aid reflect a general skew in nonprofit service provision, which in turn may reflect funding availability and the presence of lawyers engaged in pro bono services. To examine this, we compared the nationwide ISP density with that of all registered nonprofit public charities in the United States using the IRS Business Master File extract from August 2019. The correlation between the two variables was high, ranging between 0.87 and 0.92 depending on the level of geographic aggregation. In addition, in the SM we provide a regression analysis that suggests, controlling for pure demand factors such as low-income population size, that democratic vote share (likely proxying here for urban professionals) and the availability of funding (proxied by household income) are significant predictors of ISP locations. Overall, these findings suggest that both the lag hypothesis and the general skew hypothesis help explain the cluster of underserved areas.

\subsection{Optimization}

\begin{figure*}
\centering
\includegraphics[width=\linewidth]{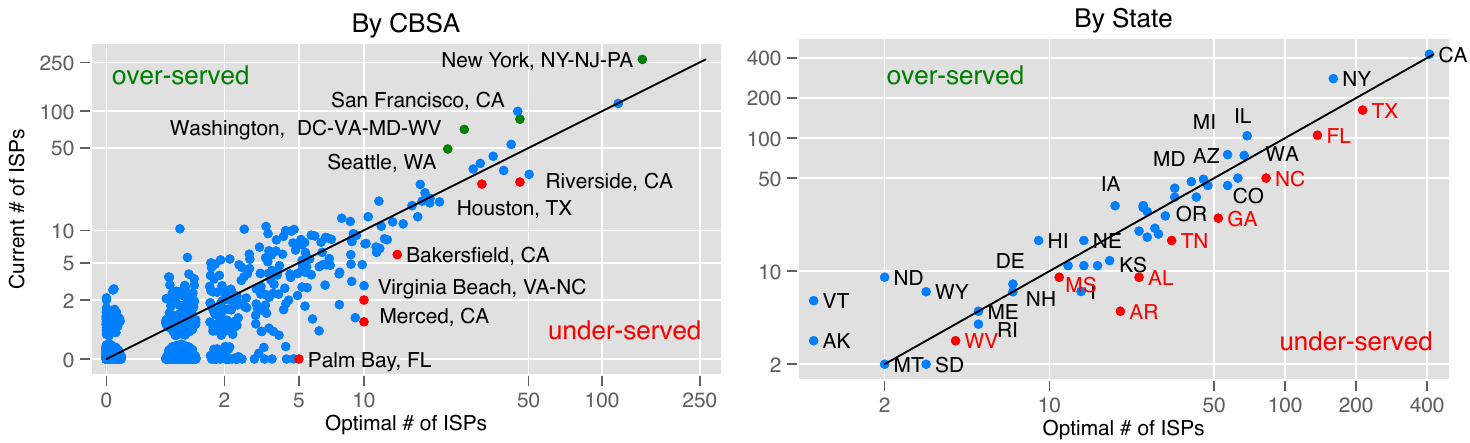}
\caption{Number of current and optimal ISPs by CBSA (left panel) and state (right) (45 degree lines in black).}
\label{fig:scatter_optimal}
\end{figure*}

Another method to identify under-served areas is to compute the optimal nationwide ISP network that maximizes service coverage for low-income immigrants and then compare the number of ISPs in the optimal and current networks. Figure \ref{fig:scatter_optimal} plots the current against the optimal number of ISPs for each CBSA (left panel) and state (right panel). Locations above (below) the 45 degree line are currently high (low) services areas relative to the optimal ISP distribution. The results suggest a clear skew in the current network. Large cities like New York City, NY, Los Angeles, CA, San Francisco, CA , Boston, MA, and Seattle, WA currently have many more ISPs than suggested by the optimal network that maximizes coverage. In contrast, Riverside, CA, Houston, TX, Phoenix, AZ, Bakersfield, CA, Virginia-Beach, VA, and Merced, CA currently have far fewer ISPs than suggested by the optimal network. The differences are similarly stark at the state level where the optimal network shifts many ISPs from relatively ovserserved states such as New York, California, Illinois, and Massachusetts to relatively underserved Southern states like Texas, North Carolina, Florida, Georgia, Tennessee, and Alabama. The differences are most striking in New York and Texas which have 119 fewer ISPs and 51 more ISPs in the optimal network compared to the current network, respectively.

\begin{figure}
\centering
\includegraphics[width=\linewidth]{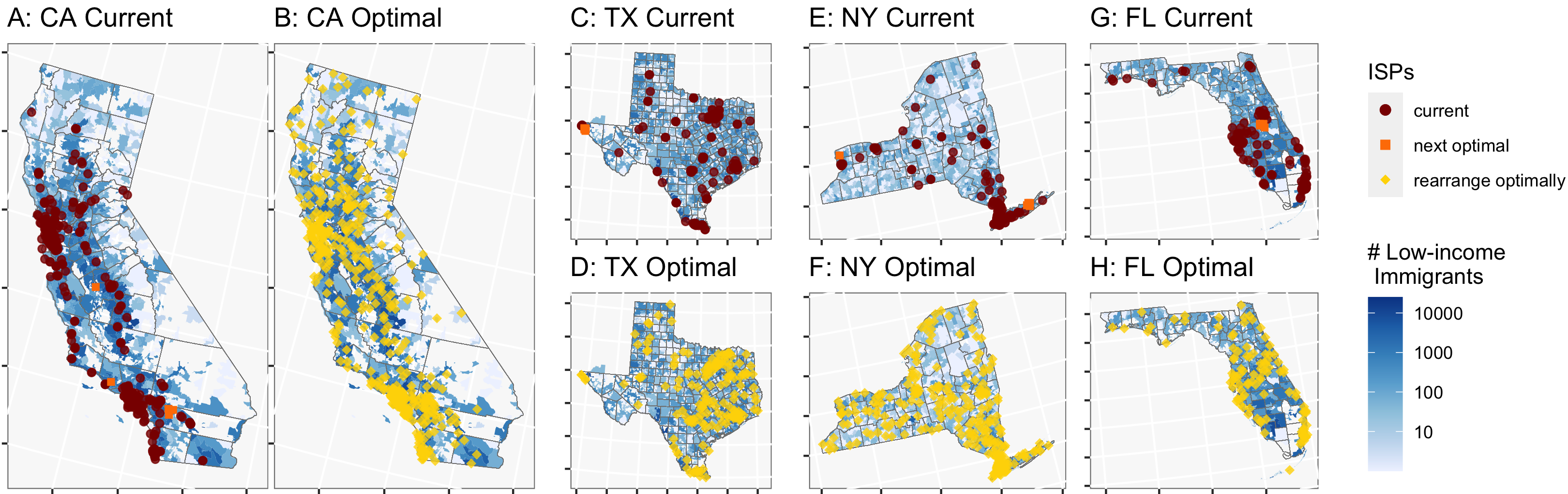}
\caption{Current (red) and optimal (yellow) ISP networks in the four largest immigrant-receiving states. The optimal locations of the next five ISPs are shown in orange.} 
\label{fig:top4}
\end{figure}

How can the current network be improved to maximize access to service? Figure \ref{fig:top4} presents the maps that results from our spatial optimization analyses in the four largest immigrant-receiving states: California, Texas, New York, and Florida, home to 57.5\% of all low-income foreign-born. For each state we show the five optimal locations to add a new ISP office (orange squares) to the existing network (red circles) to maximize access to service for the largest number of under-served low-income immigrants. For example, in California these optimal new locations are either in the area east of Los Angeles, in Santa Barbara, or in the central valley. We also show the optimal network that results from rearranging the existing ISPs in each state to maximize access to service (yellow diamonds). The optimal networks distributes many ISPs away from major urban centers to more peripheral areas and midsized cities that host many low-income immigrants. For example, in California the optimal network has many more offices distributed to the central valley area. 

\begin{figure}
\centering
\includegraphics[width=\linewidth]{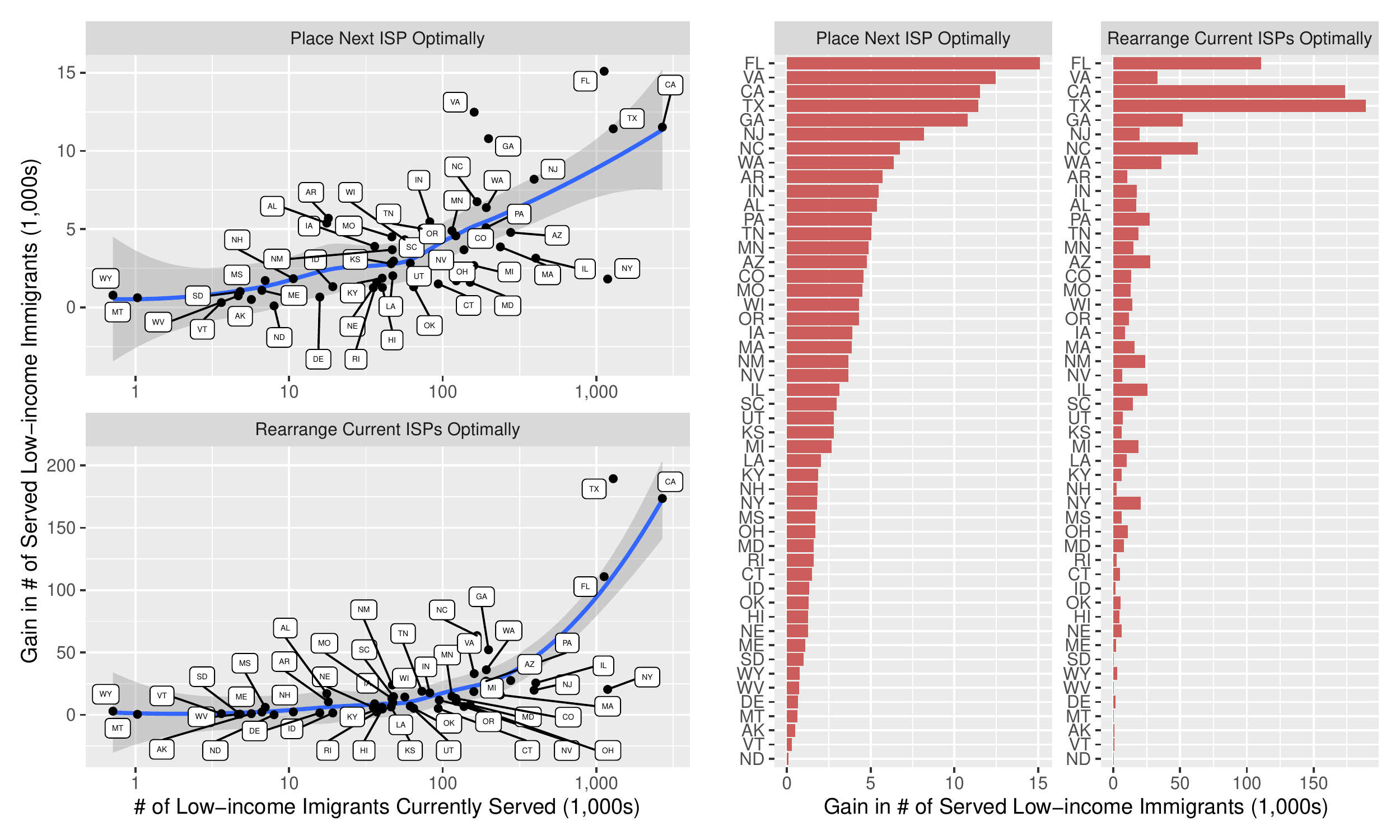}
\caption{Currently covered low-income immigrant population size and gains in coverage from adding an additional new ISP location (Panel A) and rearranging the entire ISP network optimally (B). Gains in low-income immigrant coverage from adding an additional new ISP location (C) and rearranging the entire ISP network optimally (D) by state. }
\label{fig:gains}
\end{figure}

How large are the gains from optimally restructuring the ISP networks or adding an additional ISP location in each state? Figure \ref{fig:gains} present the state-level gains in the number of low-income immigrants covered when placing an additional new ISP (top) and when restructuring the entire ISP network (bottom), plotted against the number of low-income immigrants currently served. States with larger immigrant populations experience, on average, larger gains in coverage in both types of analyses. On the one hand, New York and Illinois are well below the regression line, indicating a relatively efficient current ISP network and hence smaller gains from adding an additional new office. On the other hand, several Southern states, like Florida, Texas, Virginia and Georgia, can gain large numbers of newly covered immigrants by adding an ISP location. This pattern can be seen in the two right panels, where we show the gains by state and type of analysis. Four of the top five states in each panel are in the South, suggesting the greatest gains are in that region. The results also show considerable potential gains overall. Under the current nationwide network, about 10.4 million immigrants are covered by an ISP (i.e., they live within 12 miles of an ISP) and about 1.6 million are underserved (i.e., they live more than 12 miles from an ISP). By optimally rearranging the ISP locations within each state, 494,000 foreign-born people can gain coverage in these four states, as can 1.1 million nationwide. These gains would reduce the number of underserved low-income immigrants in those areas by about 84\% and 70\%, respectively.

\section{Discussion}

By conducting the first comprehensive study of the nationwide network of ISPs, we provide vital information to low-income immigrants, funders, and policymakers about the current state of the ISP network. By identifying underserved areas with few ISPs but many low-income immigrants, we illuminate how the current network could be improved. Our optimization analysis identifies areas where placing additional new ISP offices would yield the highest returns in terms of number of newly served low-income immigrants. Lastly, our study also contributes to the literature on civil society organizations generally and ISPs specifically. Our work highlights how in one important area of service provision---legal aid to low-income immigrants---the nonprofit sector has built a comprehensive network that filled the void left by the federal government. That said, our study also highlights how the clustering of underserved areas reflects both a general skew in non-governmental organization service provision and the fact that the ISP network has not caught up with new immigrant settlement patterns, particularly in the South. Our work also paves the way for future research. In particular, researchers could build on our work and that of \citep{chand_serving_2020} to examine the broader impact of service providers on the success of immigrants in navigating the complex legal and bureaucratic obstacles they face.
 
Our study is not without limitations. First, we cannot guarantee that our list of ISPs is complete. Yet, given our triangulation of data sources, we are confident that our list does not miss large swaths of ISPs and that the broader patterns we have identified are robust to the inclusion of those ISPs that eluded our count. Second, the ongoing COVID-19 crisis might result in shifts to the ISP network that we cannot capture in our current data. Future work could investigate how the network might change in light of the crisis.

\clearpage
\newpage
\section{Figures}

\clearpage
\newpage
\clearpage
\newpage

\clearpage
\newpage
\clearpage
\newpage

\clearpage
\bibliographystyle{aer}
\bibliography{pnas-sample}
\end{document}